\documentclass[aps,prl,reprint,showpacs,showkeys,groupedaddress]{revtex4-1}

\usepackage{amsmath, amssymb}
\usepackage{amsmath}
\usepackage{graphicx}
\usepackage{subfigure}
\usepackage{color}
\usepackage{float}
\newcommand*{\rom}[1]{}
\begin{document}


\title{Room temperature flashing Ratcheting in nano-channels}
\author{Aakash}\thanks{aakash.a@students.iiserpune.ac.in}
\author{A. Bhattacharyay}\thanks{a.bhattacharyay@iiserpune.ac.in}
\affiliation{Department of Physics, Indian Institute of Science Education and Research, Pune, Maharashtra 411008, India.}

\date{\today}

\begin{abstract}
We consider the surface-induced ratcheting transport of particles in nano-channels, particularly at room temperature. We show that at room temperature it is possible to achieve ratcheting of about 10 nm size particles in a nano-channel of about 100 nm width. The typical ratcheting velocity in such a case could be of the order of a few hundred nano-meter when the surface undulations are of a wavelength of a few hundred nano-meter and of the amplitude of a few tens of nano-meter. At room temperature, the viscosity of the fluid enabling such transport in the nano-channels comes out to be that of water. We show here a considerably large effect under realistic conditions which could be used for application in efficient filtration of particles and probably are in use in biological systems which typically work at room temperature.
\end{abstract}

\maketitle


Nanofluidics is the study of  behaviour of fluids confined to nanometer geometry (typically $\sim$ 1 - 100 nm) \cite{kirby2010micro,abgrall2009nanofluidics, eijkel2005nanofluidics}. Nanofluidics is a very active area of research at present due to the potential of huge applications. Fluids confined to these channels show transport which can be very different from that observed in bulk \cite{schoch2008transport}. Due to small channel sizes which gives control over the flow through surface fluctuation and modulations, nanofluidics finds its application where fluid samples are handled in very small quantities such as coulter counter \cite{saleh2001quantitative}, chemical and particulate separation\cite{skaug2018nanofluidic,derenyi1998ac, sonker2019separation} and analysis of bio molecules such as DNA, proteins and other ionic species \cite{esmek2019sculpturing,fu2007patterned,zeng2007self,fu2009continuous,napoli2010nanofluidic,wang2008bare,pennathur2005electrokinetic}, cell capture and counting and micropumps to name a few areas. People have applied the principle of nano-transport to perform chromatography with porous medium \cite{saridara2005chromatography,singhal2012separation,asensio2014carbon}. Nanofluidics also finds  application in nano-optics for producing tuneable microlens and for the detection of viruses and nanoparticles \cite{mitra2010nano}. Nanofluidics is also significant in the area of biotechnology and molecular diagnosis. \cite{jayamohan2013applications}. Some carbon nanotubes (CNT) can give rise to ultra-efficient transport mechanism for water and gas molecules \cite{noy2007nanofluidics,tao2018confinement,sparreboom2010transport}. Water transport rate in these nanotubes can be far exceeding than that predicted by theory. CNT (Carbon nanotubes) finds their application in desalination and purification of water. Some carbon nanotubes (CNT) exhibit remarkable electrical, thermal and mechanical properties like tensile strength etc. \cite{tans1997individual,berber2000unusually,kim2001thermal}. Apart from these there is a natural analogy between fluid flow in nanochannels and electron and holes motion in semiconductors electronic devices. And this analogy can be exploited in electronic functioning such as rectification of current and bipolar and field effect transistor actions \cite{perdigones2014correspondence}. 

\par
Though people have applied the basic ingredient of ratcheting, i.e., breaking of symmetry where they have shown that microratchet-like structures can be used to self-propel fluids, transport particles, and direct cell motion in the absence of external power sources \cite{caballero2016motion}, details about ratcheting in nano-channels remains to some extent unexplored compared to that in microfluidic systems. Moreover, ratcheting in nano-meter size channels and openings is of profound interest in connection with biological system. It has been shown by experiments that separation of species can take place by means of a ratchet device, consisting of periodically arranged triangular (ratchet) shaped obstacles \cite{bogunovic2012particle, park2017microfluidic}. 
\par
The way a paradigmatic ratcheting particle moves against a potential, the same mechanism can drive a diffusing particle in a nanofluid to move against the flow of the fluid \cite{astumian1994fluctuation,astumian1997thermodynamics,bhattacharyay2012directed,tarlie1998optimal,cisne2011particle,loutherback2009deterministic,celestino2011generic}. The very essential inversion symmetry breaking at mesoscopic scales can be realized in nanofluidic channels if the surface is elastic or semiflexible supporting oscillations so that inversion symmetry can be broken by  creating stationary modes of required wavelengths. By taking a general approach of using minimal number of Fourier modes for standing saw-tooth surface undulations on flexible surfaces in the presence of temporal oscillations, we show that surface driven ratcheting \cite{ait2003brownian,ethier2019tilted} in nano-channels particularly at room temperature can happen in water. In this work we present a detailed study of ratcheting through nano-meter size (about 100 nm radius) tubes with flexible confining surface. We show that for surface undulations of wavelength about 400 nm and amplitude about 20 nm, particles of size about 10 nm can attain a ratcheting velocity of several 100 nm per sec at room temperature. We also show that this ratcheting mechanism could be highly adjustable and velocity reversal can be achieved by tuning the phase of the surface modes. Although we focus on room temperature ratcheting in the present context, however, the same analysis can be extended to other temperatures quite readily.
\par
Surface driven ratcheting of particles in nanopores can result in efficient filtration because one can make the particles move against the flow of the fluid in the nano-channel. This would be equivalent to the ratcheting of a particle up against a potential barrier. This is where lies the importance of this phenomenon in the realm of nanofluidics by employing which one can think of a general mechanism of controlled filtration. Moreover, the existence of such a general mechanism in the realms of nanofluidics at room temperature also indicates that the mechanism might be already in use in biological systems where nanofluidcs is a norm than an exception.
\par
A recent paper by Marbach et al. \cite{marbach2018transport} reports an important general method of analyzing particle transport through nano-channels in the presence of fluctuations of the channel wall. This work develops a very essential tool of spectral analysis of motion of test particles in a nano-channel. Under lubrication approximation, this work captures diffusivity renormalisation of a test particle in a nano-channel under various surface fluctuations of the system. In the present paper, in what follows, we will be directly using the relation derived by Marbach et al. \cite{marbach2018transport}, to show that, the overdamped velocity dynamics of the test particle under surface fluctuations indicates that there exists possibility of a surface fluctuation driven ratcheting of particles in such systems.

\par
We organize the paper in the following way. For completeness we first present the salient steps of the calculations by Marbach et al. \cite{marbach2018transport} deriving the general velocity dynamics of the test particle in a nano-channel under surface fluctuations in the lubrication approximations. Then we show how this overdamped dynamics results in a ratcheting in the presence of an inversion symmetry broken patterning of the surface. We show numerical results of this ratcheting process. Following which we conclude the paper with a detailed discussion of the model.  
\section{Marginal Advection - Diffusion equation}
\par
In the following, for the sake of completeness of the presentation, we first present the basic analysis as done by Marbach et. al. in ref. \cite{marbach2018transport} to setup the stage for the analysis of ratcheting. The dynamics of probability density $p(\bf{r},\emph{t})$ of a tracer particle in the presence of velocity field $\bf{u}$($\bf{r}$,\emph{t}) is governed by Fokker-Planck equation. 

\begin{equation}
\frac{\partial p({\bf r},t)}{\partial t} = D_{0} \nabla^{2}p({\bf r},t) - \nabla\cdot[{\bf u}({\bf r},t) p({\bf r},t)] 
\end{equation}
\par
Let $h_{u}$({\bf x},\emph{t}) be the upper surface profile, where {\bf x} = ($x$, $y$) is the coordinate of the 2D plane at a fixed height $z$ and $h_{u}(\bf{x},\emph{t})$ = $H + h({\bf x},\emph{t})$. Where $H$ is the mean height of an upper fluctuating surface and $h({\bf x},t)$ is the height variation over the mean height. In this calculation \cite{marbach2018transport}, the lower surface confining the fluid layer is considered to be fixed without any loss of generality. Therefore $z$ lies between 0 and $h_{u}$({\bf x},\emph{t}). The marginal probability distribution function $p^{*} ({\bf x},t)$ is then defined as 
\begin{equation}
p^{*} ({\bf x},t) = \int_{0}^{h_{u}({\bf x},t)}  dz \: p({\bf r},t)
\end{equation}
\par
Now integrating both sides of Fokker-Planck eqn. (1) from 0 to $h_{u}$({\bf x},\emph{t}) while using the Leibnitz rule and then simplifying, one obtains an equation for marginal probability distribution
\begin{equation}
    \frac{\partial p^{*}({\bf x},t)}{\partial t} = D_{0} \nabla^{2}_{||}p^{*}({\bf x},t) - \nabla_{||}\cdot[{\bf v}({\bf x},t)p^{*}({\bf x},t)]
\end{equation}
where 
\begin{equation}
    \nabla_{||} = \hat{i}\frac{\partial}{\partial x}+\hat{j} \frac{\partial}{\partial y}
\end{equation}
and
\begin{equation}
\nabla_{||}^{2} = \frac{\partial^{2}}{\partial x^{2}} + \frac{\partial^{2}}{\partial y^{2}}.
\end{equation}
\begin{multline}
    {\bf v}({\bf x},t) = D_{0} \frac{\nabla_{||}h({\bf x},t)}{h_{u}({\bf x},t)} + \frac{1}{h_{u}({\bf x},t)} \int_{0}^{h_{u}({\bf x},t)} dz \: {\bf u_{||}} ({\bf r},t) 
\end{multline}    
is the effective velocity field that a tracer particle is getting to see. The equation (6) for effective velocity will be important for later discussions. In deriving the equation (6) one makes use of the following boundary conditions \cite{marbach2018transport}.
\par
\begin{enumerate}
\item Kinematic Boundary condition : It implies that the velocity of the fluid at the boundary is zero with respect to the boundary.
\begin{multline}
    u_{z}({\bf x},h_{u}({\bf x},t),t) - \frac{\partial}{\partial t} h_{u}({\bf x},t) - {\bf u}_{||}({\bf x},h_{u}({\bf x},t),t).\\  \nabla_{||}h_{u}({\bf x},t) = 0
\end{multline}
Where ($\nabla_{||}, \frac{\partial}{\partial z}$) and (${\bf u}_{||}, u_{z}$) denote in plane and out of plane coordinates respectively.

\item Conservation of probability at z = 0 :
\begin{equation}
    \frac{\partial}{\partial z} p({\bf x},z=0) = 0
\end{equation}

\item  Conservation of probability at $z = h_{u}({\bf x},t)$ :
\begin{equation}
    \nabla p({\bf x},z)|_{z = h_{u}({\bf x},z)}\cdot{\bf n} = 0
\end{equation}
\end{enumerate}
\par
Using equation (3) and above three boundary conditions one obtains the following equation for marginal probability density $p^{*}({\bf x},t)$
\begin{multline}
    \frac{\partial}{\partial t}p^{*}({\bf x},t) = D_{0}\nabla^{2}_{||}p^{*}({\bf x},t) - D_{0}\nabla_{||}\cdot[p({\bf x},h_{u}({\bf x},t),t)\\\nabla_{||} h_{u}({\bf x},t)] - \nabla_{||}\cdot\int_{0}^{h_{u}({\bf x},t)} dz \: {\bf u_{||}} \: ({\bf r},t) \: p({\bf r},t)
\end{multline}
\par
But this equation is not in terms of marginal probability density $p^{*}({\bf x},t)$ only. It contains terms involving $p({\bf r},t)$ also. However if one makes use of the Lubrication approximation as is shown by Marbach et al., then one can assume the equilibrium in $z$ direction and factorize the $p({\bf x},z,t)$ into $f({\bf x},t)$ and some function of $z$. Then using the normalisation one can write $p({\bf x},z,t) \approx f({\bf x},t)$. This equation greatly simplifies the above equation and one gets 
\begin{multline}
    p^{*}({\bf x},t) \approx \int_{0}^{h_{u}({\bf x},t)} dz \: f({\bf x},t) = h_{u}({\bf x},t) f({\bf x},t)
\end{multline}
\par
Therefore we have the following equation for marginal probability density $p^{*}({\bf x},t)$
\begin{multline}
    \frac{\partial}{\partial t}p^{*}({\bf x},t) = D_{0}\nabla^{2}_{||}p^{*}({\bf x},t) - D_{0} \nabla_{||}\cdot[p({\bf x},h_{u}({\bf x},t),t)\\ \nabla_{||}h_{u}({\bf x},t)] - \nabla_{||}\cdot[p^{*}({\bf x},t)\overline{{\bf u_{||}}({\bf x},t)}]
\end{multline}
Where $\overline{{\bf u_{||}}({\bf x},t)}$ is average velocity defined by :
\begin{equation}
    \overline{{\bf u_{||}}({\bf x},t)} = \frac{1}{h_{u}({\bf x},t)} \int_{0}^{h_{u}({\bf x},t)} dz \: {\bf u_{||}}({\bf x},z,t)
\end{equation}
\par
Finally the following equation is obtained for marginal probability density 
\begin{equation}
    \frac{\partial p^{*}({\bf x},t)}{\partial t} = D_{0} \nabla^{2}_{||}p^{*}({\bf x},t) - \nabla_{||}.[{\bf v}({\bf x},t)p^{*}({\bf x},t)]
\end{equation}
where 
\begin{equation}
    {\bf v}({\bf x},t) = D_{0} \frac{\nabla_{||}h({\bf x},t)}{h_{u}({\bf x},t)} + \frac{1}{h_{u}({\bf x},t)} \int_{0}^{h_{u}({\bf x},t)} dz \: {\bf u_{||}} ({\bf r},t) 
\end{equation}    
is the effective velocity field which the tracer particle is moving with.

\section{Surface induced ratcheting}
Let us have a look at how the surface fluctuations induced ratcheting can happen in nano-fluid systems. Under the lubrication approximation, the velocity of the diffusing particle as arrived at by Marbach et al., is 

\begin{equation}
    \frac{dx}{dt} = D_{0} \frac{\frac{\partial}{\partial x}h(x,t)}{H + h(x,t)} + \frac{1}{H + h(x,t)} \int_{0}^{H+h(x,t)}dz \: u(x,z,t)
\end{equation}
where we have considered only one of the planar coordinates ($x$,$y$) for the sake of simplicity.
In the above equation, there are two terms on the RHS of which the first term gives the average velocity induced by diffusion and surface gradients and the second term stands for the advection of the particle by the fluid of velocity $u(x,z,t)$. If the first term can produce a velocity which is  opposite in direction to the second term and can overcome the advection then the particle can actually move in opposite direction to the fluid flow resulting in enhanced filtration. As it is obvious from the presence of gradient term $\frac{\partial}{\partial x}h(x,t)$, the sign of the velocity due to this first term can be reversed and we can always get this part of velocity to be opposite to that coming from the second term. However, this will require a breaking of inversion symmetry at smaller scales.
\par
Thus in the presence of diffusive coupling along with gaussian white noise with zero mean and unit variance position of the particle obeys the following Stochastic differential equation :

\begin{equation}
    \frac{dx}{dt} = \frac{D_{0}}{H} \frac{\partial}{\partial x}h(x,t) + \sqrt{2D_0} \:\xi (t)
\end{equation}
Where we have gone by the assumption $H \gg h(x,t)$ which is the lubrication approximation. $\xi$(t) is gaussian white noise with zero mean, unit variance and delta correlations. This equation is the overdamped dynamics of a particle in the potential $-\frac{D_0}{H} h(x,t)$. We model the surface undulations $h(x,t)$ by the superposition of two travelling waves moving in opposite direction with frequency $\omega$ and wave-number $k$.
\par
\begin{multline}
    h(x,t) = h_0\: sin(\omega_0 t + \phi)\\ \left[ \sum_{n = 1}^{2} (-1)^n \frac{sin(\omega t - nkx)}{n}  - \sum_{n=1}^{2} (-1)^n \frac{sin(\omega t + nkx)}{n}  \right] 
\end{multline}
Each of the individual wave is nothing but the two term truncation of a Fourier series of a saw-tooth profile. First term of the $h(x,t)$ provides a global oscillation with frequency $\omega_0$ and phase $\phi$. This particular $h(x,t)$ profile provides the essential ingredient of ratcheting, i.e., breaking of inversion symmetry. In FIG.(1) we have plotted the potential profile showing broken symmetry obtained from $h(x,t)$ profile.

\begin{figure}[ht]
  \includegraphics[width=9.0cm,height=6.1cm]{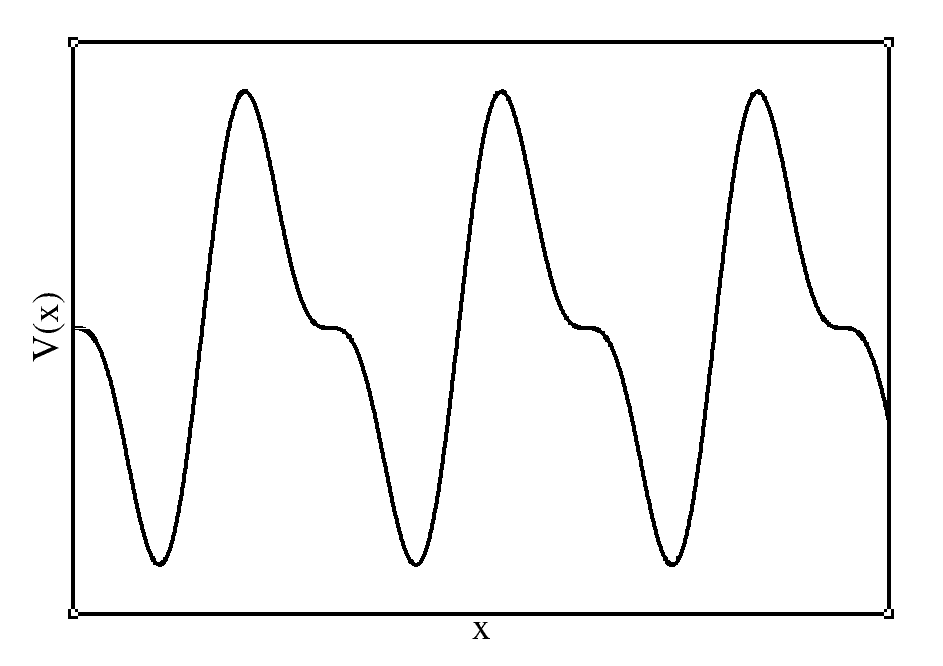}
\caption{Ratcheting potential showing the broken inversion symmetry}
\label{fig:5}       
\end{figure}
\par
After simplifying the above expression for $h(x,t)$ using trigonometric identities one can obtain the following expression for $h(x,t)$ :

\begin{multline}
    h(x,t) = h_0\: sin(\omega_0 t + \phi)\\ \left[ 2 \;cos(\omega t) \;sin(kx) - cos(\omega t) \;sin(2kx) \right] 
\end{multline}

\par
After substituting above expression for $h(x,t)$ in equation (17) on can obtain the following Stochastic equation of motion for particle :

\begin{multline}
    \frac{dx}{dt} = 2kD_{0}\frac{h_0}{H}\: sin(\omega_0 t + \phi)\:cos(\omega t)\\\left[ cos(kx) - cos(2kx) \right] + \sqrt{2D_0} \:\xi (t)
\end{multline}

\par
Now we introduce the dimensionless length $\bar{x}$ defined by $x =\lambda \bar{x} $, where $\lambda = \frac{2\pi}{k}$. In terms of this dimensionless length we get the following equation of motion which will  be our working equation.

\begin{multline}
    \frac{d\bar{x}}{dt} = 4 \pi \bar{D}_{0}\frac{h_0}{H}\: sin(\omega_0 t + \phi)\:cos(\omega t)\\\left[ cos(2\pi\bar{x}) - cos(4\pi \bar{x}) \right] + \sqrt{2\bar{D}_0} \:\xi (t)
\end{multline}

\par
Here $\bar{D}_0 = \frac{D_0}{\lambda^2} $ and has the dimensions of inverse time. Thus inverse of $D_0 $ provides a time scale which we will use later on to analyse our results. This equation (21) will be our working equation for numerical simulation.

\section{Numerical Results}

\par
In FIG.(2) we show the ratcheting trajectories as obtained from the direct simulation of eqn.(21). Here we have plotted x vs t for $\bar{D}_0$ = 30, 35, 60 and 90 $s^{-1}$, $h_0/H = 0.2$ (i.e. considering $h_0$ = 20 nm and H = 100 nm),$\; \omega_0 = \omega = 50$ rad s$^{-1}$ and $ \phi = 1.4 $ rad. The time step for simulation is $\Delta t $ = 0.001 s and we have employed Euler-Maruyama method for numerical integration. We have calculated velocity of Brownian particles by taking ensemble average over 20000 realisations. The typical ratcheting velocity which we obtain is 0.1 - 0.8 $\lambda$ per sec. 
\begin{figure}[ht]
  \includegraphics[width=9.0cm,height=6.4cm]{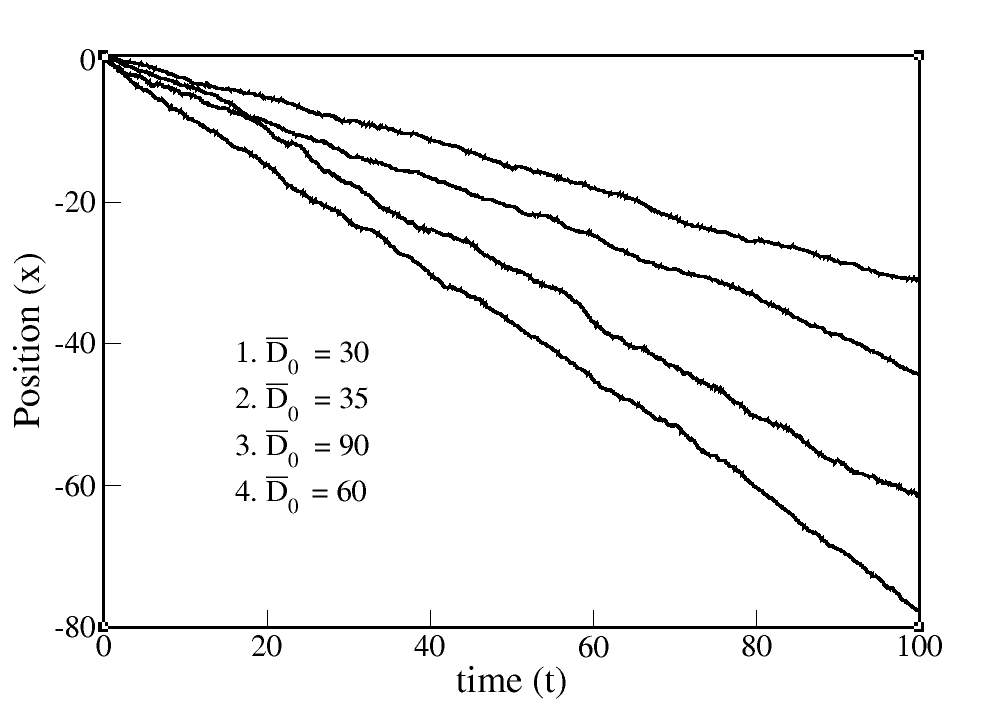}
\caption{Position of particle showing directed motion (x is in the units of $\lambda$ and t is in sec., $h_0/H = 0.2, \omega = \omega_0 = 50\; rad/s, \phi = 1.4 \;rad$)}
\label{fig:5}       
\end{figure}
\par
Since our working equation (21) was independent of $\lambda$ after scaling, one can increase(decrease) $\lambda$ by the corresponding increase(decrease) in the value of $D_0$ keeping $\bar{D}_0 = D_0/\lambda^2$ constant. To interpret our results in terms of the room temperature ($\sim$ 300 K), we will make use of the Stokes-Einstein relation in what follows. Since, ratcheting is a weakly non-equilibrium phenomenon essentially in the slow dynamics regime, making use of Stokes-Einstein relation is appropriate in this context. The Stokes-Einstein relation is
\begin{equation}
   D_0 = \frac{k_B T}{6 \pi \nu r},
\end{equation}
Which gives the diffusion constant of the Brownian particle in terms of the temperature $T$, the radius of the particle $r$ and the viscosity of the medium $\nu$. In the Stokes-Einstein relation $k_B$ is Boltzmann constant. One can estimate the temperature of the medium over which this ratcheting happens corresponding to a reasonably fixed $r$ and $\nu$ for a given $D_0$. Similarly by the same argument one can see that for a given particle size one can get the whole continuous range of temperature T for which ratcheting will happen at different $\lambda$ and, of course, room temperature is just a particular case. For example, substituting the typical viscosity, $\nu$  = $2\times 10^{-3}$ Pa s and a typical radius of the Brownian particle to be 10 nm = $10^{-8} \;m$ for $\bar{D}_0 = 70 \;s^{-1} $ which corresponds to $D_0$ = 1.75 $\times 10^{-11} m^2/s$ by choosing $\lambda\; \approx $  400.0 nm we get T $\approx$ 300 K.
\par
We can see from FIG.(3) that the magnitude of ratcheting velocity first increases with the increase in the $\Bar{D}_0$ attains a maximum and then finally decreases. The maximum of $\Bar{v}$ occurs at $\Bar{D}_0 \sim 70 \;s^{-1}$. We have studied the effect of global phase $\phi$ on the ratcheting velocity $\Bar{v}$ which shows a periodic change in the ratcheting velocity as shown in FIG.(4) changing from negative to positive, i.e., direction reversal of the average motion due to ratcheting of the Brownian particle can be achieved by adjusting the phase of standing waves on the surface. 
\begin{figure}[ht]
  \includegraphics[width=9.0cm,height=6.4cm]{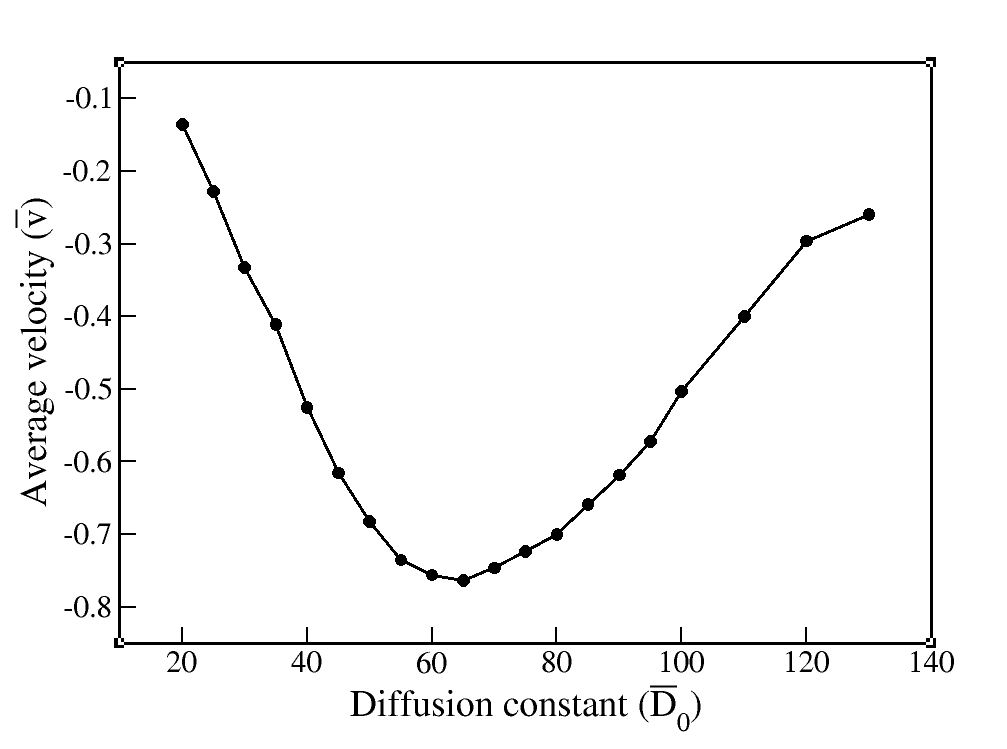}
\caption{Variation of Average velocity with Diffusion constant $\Bar{D}_0$($\Bar{v}$ is in the unit of $s^{-1}$ and $\Bar{D}_0$ is in the unit of $s^{-1}$, $h_0/H = 0.2, \omega = \omega_0 = 50\; rad/s, \phi = 1.4 \;rad$)}
\label{fig:5}       
\end{figure}
\begin{figure}[H]

\includegraphics[width=9.0cm,height=6.4cm]{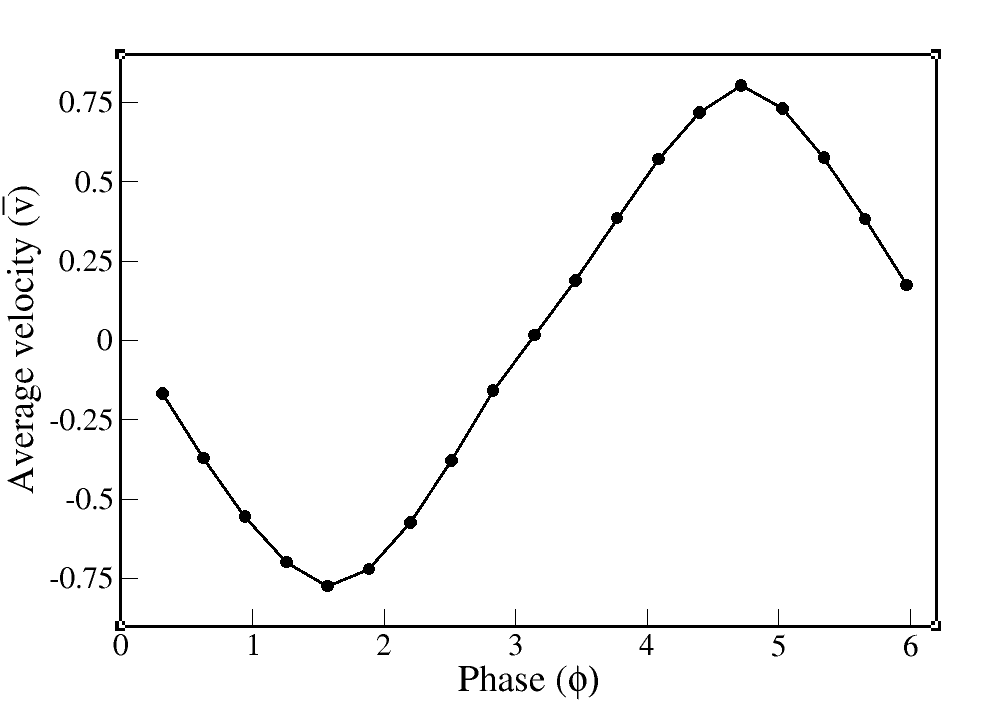}
\caption{Variation of Average velocity with phase $\phi \:$($\Bar{v}$ is in the unit of $s^{-1}$ and $\phi$ is in the units of radian, $h_0/H = 0.2, \omega = \omega_0 = 50\; rad/s, \bar{D}_0 = 70 \;s^{-1}$)}
\label{fig:6}       
\end{figure}

\par
The size of surface undulations compared to the nano-tube diameter plays important role in ratcheting. One can see from FIG.(5) that magnitude of ratcheting velocity increases on increasing the $h_0/H$ however the limitation to the $h_0/H$ scale comes in the present model from the lubrication approximation. It is because of the underlying lubrication approximation that we have kept $h_0/H $ between 0.2 to 0.3 and have not gone to higher values, however, this is a very reasonable range of the amplitude of undulation of the surface confining the nano-fluid for the phenomenon of ratcheting.
\par
We have studied the effect of frequency on the ratcheting velocity and we found that ratcheting velocity remains practically unaffected by the change in frequency of driving (shown in FIG.(6)). There exists, however, a certain range of frequencies below and over which ratcheting no longer happen for the typical values of the other parameters used. We have taken the frequency of surface undulations to be 50 rad/sec giving a time period of the order of 0.1 sec for surface undulations. This also indicates that, our choice of the discrete time scale of $10^{-3}$ sec in the simulations is good enough.

\begin{figure}[ht]
\includegraphics[width=9.0cm,height=6.5cm]{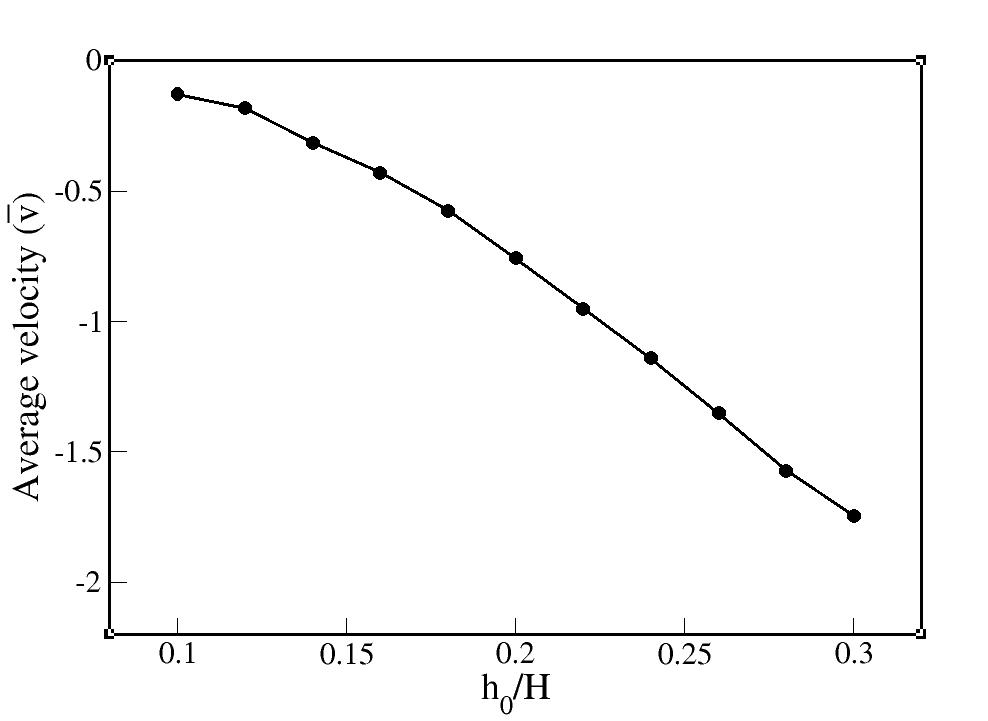}
\caption{Variation of Average velocity with $h_0/H\:$($\Bar{v}$ is in the unit of $s^{-1}$, $\bar{D}_0 = 70\;s^{-1}, \omega = \omega_0 = 50\; rad/s, \phi = 1.4 \;rad$) }
\label{fig:7}       
\end{figure}

\begin{figure}[ht]
  \includegraphics[width=9.0cm,height=6.4cm]{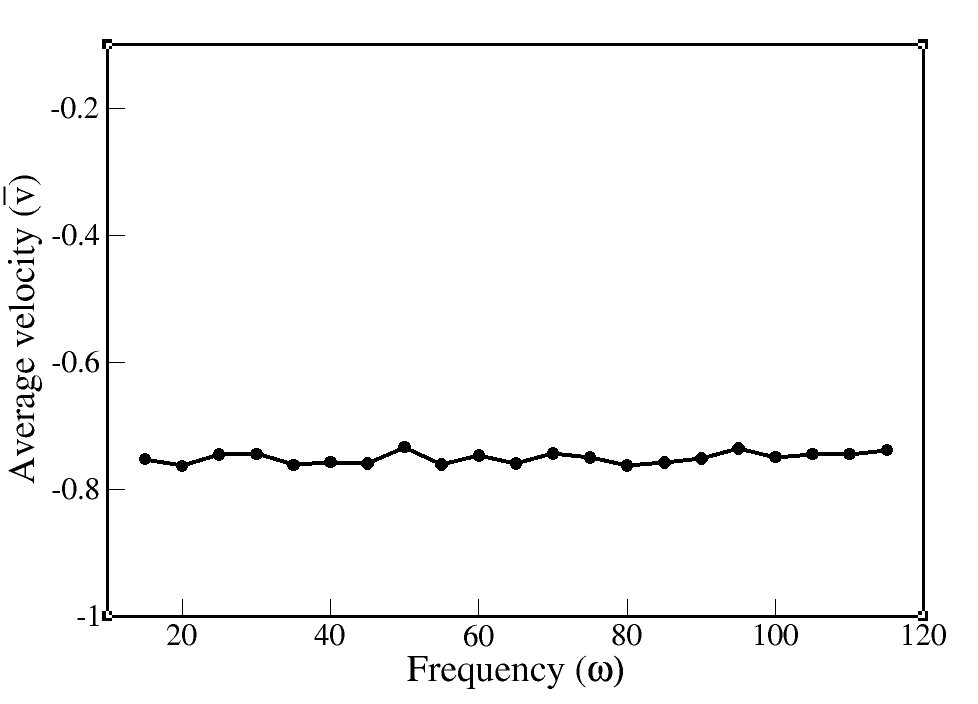}
\caption{Variation of average velocity with frequency of oscillation ($\bar{v}$ is in unit of s$^{-1}$  and  $\omega$ is in rad $s^{-1}$, $h_0/H = 0.2, \bar{D}_0 = 70 \;s^{-1}, \phi = 1.4 \;rad$)}
\label{fig:5}       
\end{figure}
\par
We define the efficiency $\eta$ of ratcheting by the ratio of mean square velocity of the Brownian particle and square of the maximum particle velocity of undulation wave on the surface. Note that, the energy is spent mostly by an external agent to create the surface undulations and that is why we define the efficiency in this way. The thermal energy essentially assists in the process of ratcheting by providing the required jitters without contributing much energetically as is done by the non-equilibrium forcing.
\begin{equation}
    \eta = \frac{\bar{v}^2 \lambda^2}{\omega^2 h_0^2}
\end{equation}
\par
We study the variation of efficiency, $\eta$ with diffusion constant $\bar{D}_0$, viscosity $\nu$, wavelength $\lambda$ of surface undulations and the size of Brownian particle r. In FIG.(7) we have plotted the variation of efficiency($\eta$) with diffusion constant($\bar{D}_0$). By Stokes-Einstein relation ($\lambda^2 \bar{D}_0 = k_B T/6 \pi \nu r$), we see that if we fix $\lambda$, $\nu$ and r then $\bar{D}_0 \propto T$. We have fixed $\lambda$ = 400 nm, $\nu$ = 2 $\times10^{-3}$ Pa s and r = 10 nm. We find by numerical simulations that efficiency of ratcheting increases with the increase in diffusion constant (or increase in temperature T), attains a maxima at $\bar{D}_0 \sim 55 \;s^{-1}$(which corresponds to T $\sim$ 240 K) and then start falling as shown in FIG.(7). The value of the temperature mentioned to be $T\sim 240$ K is dependent on the choice of the wavelength $\lambda$ which has been fixed here to 400 nm. We need to fix the $\lambda$ to get the velocity scale. Thus, just by adjusting the $\lambda$ a little one can also interpret this graph for the exact room temperature as well.
\begin{figure}[ht]
\includegraphics[width=9.0cm,height=6.4cm]{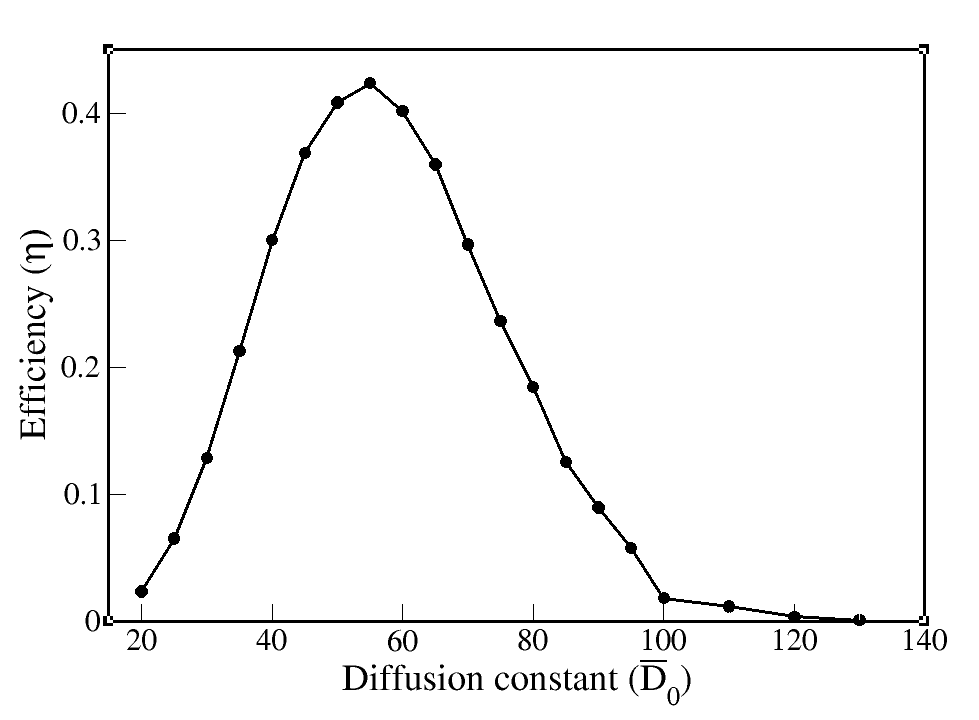}
\caption{Variation of efficiency $\eta$ with diffusion constant $\bar{D}_0$ ($\bar{D}_0$ is in the units of $s^{-1}$, $h_0/H = 0.3, \omega = \omega_0 = 50\; rad/s, \phi = 1.4 \;rad$, $\lambda$ = 400 nm, $\nu$ = 2 $\times\;10^{-3}$ Pa s, r = 10 nm)}
\label{fig:5}       
\end{figure}

In FIG.(8) we have plotted the variation of efficiency($\eta$) with wavelength($\lambda$) of surface modes. Since $\lambda^2 \bar{D}_0 = k_B T/6\pi \nu r$, therefore if we fix $\nu$, r and T then $\lambda^2 \propto 1/\bar{D}_0$. Keeping the viscosity($\nu$ = 2$\times 10^{-3}$ Pa s), particle size(r = 10 nm) and temperature(T = 300 K) fixed we find that efficiency first increases with wavelength, attains a maximum and then starts falling. This result is particularly important because for given particle size and the nature of the confined fluid at room temperature it tells us of the wavelength selection for the surface undulations such that the efficiency is maximum.
\begin{figure}[ht]
  \includegraphics[width=9.0cm,height=6.4cm]{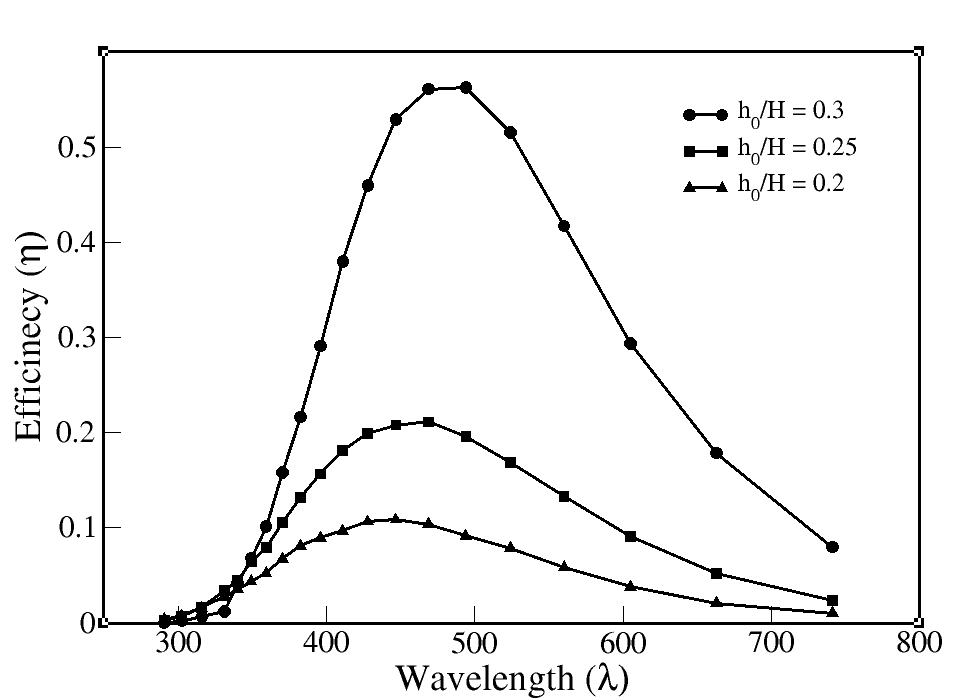}
\caption{Variation of efficiency $\eta$ with wavelength ($\lambda$) of surface undulations  ($\lambda$ is in the units of nm, $\nu$ = 2 $\times\;10^{-3}$ Pa s, r = 10 nm, T = 300 K)}
\label{fig:5}       
\end{figure}
\par
In FIG.(9) we have plotted the variation of efficiency with viscosity($\nu$) of surrounding medium. Since $\lambda^2 \bar{D}_0 = k_B T/6\pi \nu r$, thus keeping $\lambda$, r and T constant $\bar{D}_0 \propto 1/\nu$.  We have fixed $\lambda$ at 400 nm, particle radius,r at 10 nm and T at 300 K. We find that efficiency first increases with increase in viscosity, attains a maximum and then starts falling. Particularly important is the fact that, the efficiency peaks at a viscosity which is close to that of water at room temperature. 
\begin{figure}[ht]
  \includegraphics[width=9.0cm,height=6.4cm]{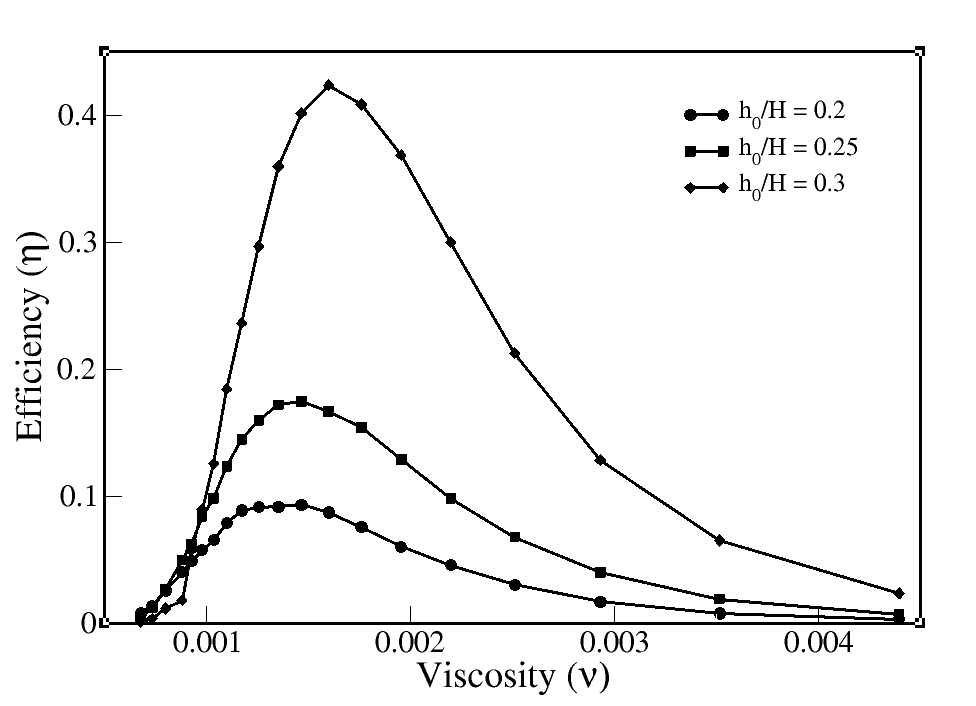}
\caption{Variation of efficiency $\eta$ with viscosity $\nu$ ($\nu$ is in the units of Pa-s, $\lambda$ = 400 nm, r = 10 nm, T = 300 K)}
\label{fig:5}       
\end{figure}
\par
Finally in FIG.(10) we study the variation of efficiency($\eta$) with particle radius(r). Since $\lambda^2 \bar{D}_0 = k_B T/6\pi \nu r$, thus keeping $\lambda$, $\nu$ and T constant $\bar{D}_0 \propto 1/r$. We have fixed wavelength,$\lambda$ = 400 nm, $\nu$ = 2$\times10^{-3}$ Pa s and T = 300 K. We find that efficiency first increases with increase  in  particle radius,  attains  a  maxima  and  then  starts falling.
\begin{figure}[ht]
  \includegraphics[width=9.0cm,height=6.4cm]{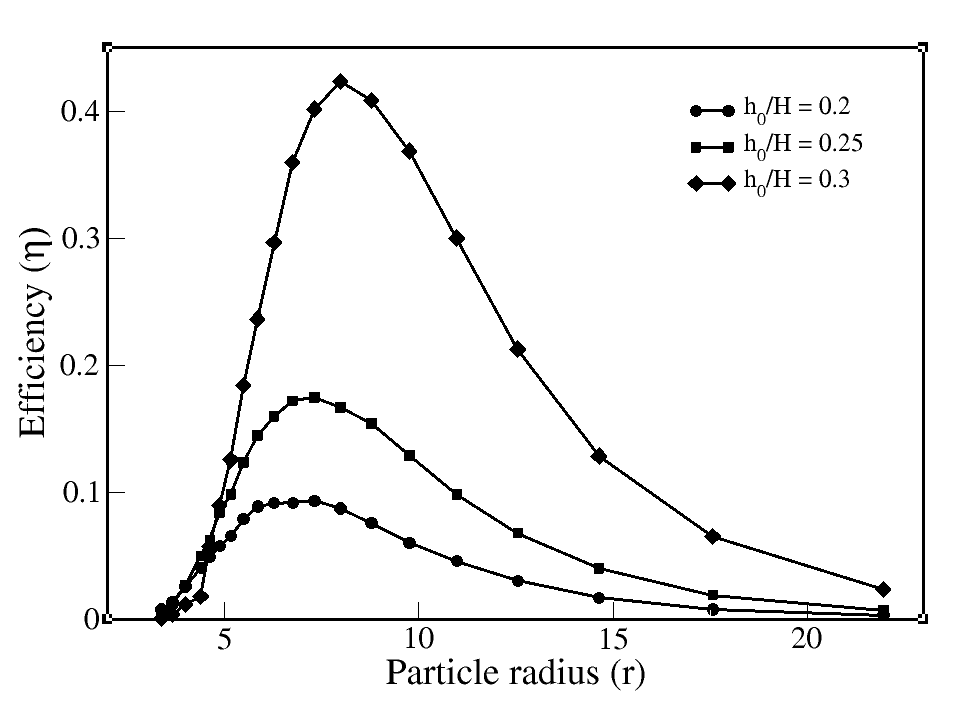}
\caption{Variation of efficiency $\eta$ with particle size $r$ (r is in nm, $\lambda$ = 400 nm, $\nu$ = 2 $\times\;10^{-3}$ Pa s,T = 300 K)}
\label{fig:5}       
\end{figure}
\par
Particular points to note from these graphs are (a) The value of the viscosity at which the efficiency peaks is quite close to that of water at room temperature, (b) The radius of the particle where the efficiency peaks being about 10 nm is quite consistent with the tube width of about 100 nm and undulation amplitude of a few tens of nano-meter, (c) The efficiency falls quite rapidly for particle size less than about 5 nm at room temperature. This indicates the fine tuning needed in the scales of the intermittent driving and the temperature for ratcheting and that is quite well known. These plots are clearly revealing that, at room temperature, there exists a window of ratcheting transport in water of about 10 nm size particles while driven by surface modes of wavelength of a few hundred nano-meter and amplitude of oscillation being a few tens of nano-meter.  
\section{discussion}
\par
We have shown in this paper that, based on a broader framework  as  established by Marbach et  al.,\cite{marbach2018transport} there can happen surface  fluctuations  driven  ratcheting of Brownian particles in a nanofluidic channel in the presence of a medium of water.  This ratcheting of particles can result in velocity of particles against the advective motion of the particles in the fluid as long as lubrication approximation is satisfied. This in principle can result in effective filtration of particles which are smaller than the pore size. The model we present here for the room temperature scenario in particular, however, the same method can easily be generalized to other fluids and temperatures as well.
\par
The main result of this work is that ratcheting velocity turns out to be of the order of $\sim \lambda$ which in original units transform into about 500 nm/sec (for maximum efficiency) which is a good number at room temperature. It turns out that, water is the most suitable fluid for ratcheting at room temperature in the nano-channels. Further this model explain the ratcheting of a whole range of particle sizes as shown in FIG.(10). This is a remarkable effect that the boundary undulations can produce ratcheting for a very large spectrum of particle sizes typical of nano-fluidic systems. 
\par 
This  mechanism of ratcheting which is possible in nano-fluids under surface fluctuations is a generic one. Most probably, such a mechanism  could be in use in some biological systems where particle transport happens without any directional drive. Effects of other finer details on such transport can be addressed once the existence of the basic mechanism is established by experiments. We hope that, possibility of the application of this mechanism in designing efficient filters is quite possible given the ability of nano-scale engineering of such channels these days.
\section{ACKNOWLEDGEMENTS}
Aakash would like to thank the Council of Scientific and Industrial Research (CSIR), India for providing funding during this research.
\bibliography{references.bib}
\end{document}